\documentclass[12pt]{article}
\usepackage{epsfig}
\usepackage{multirow}

\newcommand{\mysection}{\setcounter{equation}{0}\section}

\def\beq{\begin{equation}}
\def\eeq{\end{equation}}
\def\beqa{\begin{eqnarray}}
\def\eeqa{\end{eqnarray}}
 
\newlength{\dinwidth} \newlength{\dinmargin}
\begin{document}

\begin{center}
{\Large \bf Single-top transverse-momentum distributions at approximate NNLO}
\end{center}
\vspace{2mm}
\begin{center}
{\large Nikolaos Kidonakis}\\
\vspace{2mm}
{\it Department of Physics, Kennesaw State University,\\
Kennesaw, GA 30144, USA}
\end{center}
 
\begin{abstract}
I present approximate next-to-next-to-leading-order (aNNLO)  
transverse-momentum ($p_T$) distributions in single-top production processes. 
The aNNLO results are derived from next-to-next-to-leading-logarithm (NNLL)  
resummation of soft-gluon corrections in the differential cross section.  
Single-top as well as single-antitop $p_T$ distributions are shown in 
$t$-channel, $s$-channel, and $tW$ production for LHC energies.
\end{abstract}

\mysection{Introduction}

The study of top quarks is a central part of the current collider programs.
While the main production mode at the LHC is top-antitop pair production, 
single-top production is an important set of processes which have been observed 
at the Tevatron and the LHC.

The production of single tops or single antitops can proceed via three different partonic-channel processes. The numerically dominant one is the $t$-channel, which involves the exchange of a space-like $W$ boson, i.e. processes of the form $qb \rightarrow q' t$ and ${\bar q} b \rightarrow {\bar q}' t$ for single-top production, as well as $q {\bar b} \rightarrow q' {\bar t}$ and ${\bar q} {\bar b} \rightarrow {\bar q}' {\bar t}$ for single-antitop production. 
The numerically smallest single-top process at the LHC is the $s$-channel, 
which involves the exchange of a time-like $W$ boson, i.e. processes of the 
form $q{\bar q'} \rightarrow {\bar b}t$ for single-top production and 
$q{\bar q'} \rightarrow b {\bar t}$ for single-antitop production. 
The third channel is associated $tW$ production, via the processes 
$bg \rightarrow tW^-$  and ${\bar b} g \rightarrow {\bar t} W^+$, which
is quite significant at the LHC. 

The complete next-to-leading order (NLO) corrections to the 
differential cross section for $t$-channel and $s$-channel production appeared 
in Ref. \cite{BWHL} and for $tW$ production in Ref. \cite{SHZ}. 
Further results for the NLO top transverse-momentum, $p_T$, distributions at LHC energies in $t$-channel production have appeared in \cite{CFMT,SYMCtch,FGMStch,FRTtch,PFtch}; for $s$-channel production in \cite{HCSYsch}; and for $tW$ production in \cite{RetW}. 

Higher-order corrections (beyond NLO) can be calculated from soft-gluon 
resummation. Such calculations for single-top production appeared first at 
next-to-leading-logarithm (NLL) accuracy in Ref. \cite{NKtevlhc}. 
More recently, calculations were performed at next-to-next-to-leading-logarithm (NNLL) accuracy in the resummation in Ref. \cite{NK}. 
Approximate next-to-next-to-leading-order (aNNLO) corrections were calculated from the NNLL result for $t$-channel, $s$-channel, and 
$tW$ production \cite{NK}. As shown and discussed in detail in 
\cite{NKtevlhc,NK}, the soft-gluon corrections numerically 
dominate the cross section, and thus the soft-gluon approximation works very 
well. The approximate and exact NLO cross sections for single-top production 
in all channels are within a few percent of each other for all LHC and Tevatron energies, and for the $t$-channel this is also known to hold at NNLO since the recent results in \cite{tchNNLO}. 

The resummation in \cite{NKtevlhc,NK} is performed 
for the double-differential cross section, and thus it enables the calculation of differential distributions in addition to total cross sections. The transverse-momentum distribution of the top or antitop is very interesting because effects due to new physics may appear at large $p_T$. 
The calculation of these $p_T$ distributions in all three single-top channels 
at LHC energies is the topic of this paper. Results for the $t$-channel $p_T$ 
distributions have been published before in \cite{NKtchpt}, so here we update them and give new results for the new 13 TeV LHC energy. For the $s$-channel and $tW$ production we provide new results.

Our work follows the standard moment-space perturbative QCD resummation formalism. Results for $t$-channel $p_T$ distributions based on another approach, soft-collinear effective theory (SCET), have appeared in \cite{WLZ}, and the differences between the moment-space and SCET approaches have been described in \cite{NK}.

In the next section I describe the kinematics and give some details for the calculation of the aNNLO corrections. We present numerical results for the single-top and single-antitop $p_T$ distributions in the $t$-channel in Section 3, in the $s$-channel in Section 4, and in the $tW$-channel in Section 5. We conclude in Section 6. 

\mysection{Kinematics for aNNLO single-top distributions}

We study single-top production in collisions of protons $A$ and $B$ 
with momenta $p_A+p_B \rightarrow p_3+p_4$. The hadronic kinematical 
variables are $S=(p_A+p_B)^2$, $T=(p_A-p_3)^2$, and $U=(p_B-p_3)^2$.
The partonic reactions have momenta $p_1+p_2 \rightarrow p_3+p_4$.
The partonic kinematical variables are 
$s=(p_1+p_2)^2$, $t=(p_1-p_3)^2$, and $u=(p_2-p_3)^2$, 
with $p_1=x_1 p_A$ and $p_2=x_2 p_B$. 
We also define the threshold variable $s_4=s+t+u-m_3^2-m_4^2$. If we denote the top-quark mass by $m_t$ and the $W$-boson mass by $m_W$, then for $t$-channel and $s$-channel production $m_3=0$ and $m_4=m_t$, while for $tW$ production $m_3=m_t$ and $m_4=m_W$. We note that $s_4$ vanishes at partonic threshold for each process.

The resummation of soft-gluon corrections follows from the factorization of the 
differential cross section into hard, soft, and jet functions in the partonic 
processes \cite{NKtevlhc,NK}. The resummed result is then used 
to generate approximate higher-order corrections.
The soft-gluon corrections have the form of logarithmic plus distributions, $[\ln^k(s_4/m_t^2)/s_4]_+$, where $0 \le k \le 2n-1$ for the $n$th order perturbative QCD corrections. The approximate NNLO soft-gluon corrections to the double-differential partonic cross section, $d^2{\hat \sigma}/(dt \,du)$, are of the form 
\beq 
\frac{d^2{\hat \sigma}^{(2)}}{dt \, du}=F_{\rm LO} \frac{\alpha_s^2}{\pi^2} 
\sum_{k=0}^3 C_k^{(2)} \left[\frac{\ln^k(s_4/m_t^2)}{s_4}\right]_+
\label{2corr}
\eeq
where $\alpha_s$ is the strong coupling, and $F_{\rm LO}$ denotes the leading-order (LO) contributions, i.e. $d^2{\hat \sigma}^{(0)}/(dt \, du)=F_{\rm LO} \, \delta(s_4)$. 
The aNNLO coefficients $C_k^{(2)}$ are in general different for each partonic process. 
The leading coefficient, $C_3^{(2)}$, depends only on color factors and it equals 3$C_F$ for $t$-channel and $s$-channel production, and $2(C_F+C_A)$ for $tW$ production, where $C_F=(N_c^2-1)/(2N_c)$ and $C_A=N_c$, with $N_c=3$ the number of colors.
The subleading coefficients $C_2^{(2)}$, $C_1^{(2)}$, and $C_0^{(2)}$
are in general functions of $s$, $t$, $u$, $m_t$, and the factorization scale $\mu_F$, and (for $C_1^{(2)}$ and $C_0^{(2)}$) also the renormalization scale $\mu_R$. These coefficients have been determined from two-loop calculations for all partonic processes contributing to these channels \cite{NKtevlhc,NK}. NLL resummation \cite{NKtevlhc} is sufficient to calculate all aNNLO coefficients except $C_0^{(2)}$, which is fully determined only by NNLL resummation \cite{NK}. The one-loop soft-anomalous dimension \cite{NKtevlhc,NK} for each process contributes to all subleading coefficients, while the two-loop soft-anomalous dimension \cite{NK} for each process contributes to $C_0^{(2)}$.

For the $t$-channel processes $qb \rightarrow q't$  the LO terms are
\beq
F_{\rm LO}^{qb \rightarrow q't}= 
\frac{\pi \alpha^2 V_{tb}^2 V_{qq'}^2}{\sin^4\theta_W}
\frac{(s-m_t^2)}{4s(t-m_W^2)^2} \, .
\eeq
Here $\alpha=e^2/(4\pi)$, $V_{ij}$ are elements of the CKM matrix, and $\theta_W$ is the weak mixing angle.

For the $t$-channel processes ${\bar q}b \rightarrow {\bar q'} t$ the LO terms are 
\beq
F_{\rm LO}^{{\bar q}b \rightarrow {\bar q'}t}= 
\frac{\pi \alpha^2 V_{tb}^2 V_{qq'}^2}{\sin^4\theta_W}
\frac{[(s+t)^2-(s+t)m_t^2]}{4s^2(t-m_W^2)^2} \, .
\eeq

For the $s$-channel processes $q {\bar q'} \rightarrow {\bar b} t$ the LO terms are
\beq
F_{\rm LO}^{q {\bar q'} \rightarrow {\bar b} t}= 
\frac{\pi \alpha^2 V_{tb}^2 V_{qq'}^2}{\sin^4\theta_W} 
\frac{t(t-m_t^2)}{4s^2(s-m_W^2)^2} \, .
\eeq

For the associated production process $bg \rightarrow tW^-$ the LO terms are
\beq
F_{\rm LO}^{bg \rightarrow tW^-}= 
\frac{\pi V_{tb}^2 \alpha_s \alpha}{12 m_W^2 \sin^2\theta_W s^2}
\left(\frac{A_1}{(u-m_t^2)^2}-\frac{2 A_2}{(u-m_t^2) s}+\frac{2 A_3}{s^2}\right) \, ,
\eeq
where
$A_1=-(u-m_W^2)(s-m_t^2-m_W^2)(2m_W^2+m_t^2)/2
-(t-m_t^2)(-2m_W^4+m_W^2 m_t^2+m_t^4)/2
-2 (u-m_W^2) m_t^2 (2 m_W^2+m_t^2)$; 
$A_2=-(t-m_t^2)(-m_W^2+m_t^2)m_W^2-(u-m_W^2) (t-m_t^2) m_t^2/2
-(u-m_t^2) (u-m_W^2) m_t^2 / 2 - s m_W^2 m_t^2-s m_t^4/2$; and 
$A_3=-s (u-m_t^2) (2 m_W^2+m_t^2)/4$.

To calculate the hadronic differential cross section we convolute the partonic cross section with parton distribution functions (pdf). 
We use the MSTW2008 NNLO pdf \cite{MSTW} in our numerical results below unless otherwise noted. We do that for consistency with the results shown in our previous work \cite{NK}. But we also discuss results using the recent MMHT 2014 NNLO pdf \cite{MMHT}. As we will see, the shape of the distributions is not affected by this choice of pdf for any of our results.

The transverse-momentum distribution of the top quark (or of the antitop) is given by
\beqa
\frac{d\sigma}{dp_T}&=&
2 \, p_T \int_{Y^-}^{Y^+} dY 
\int_{x_2^-}^1 dx_2 
\int_0^{s_{4max}} ds_4 \, 
\frac{x_1 x_2 \, S}{x_2 S+T_1} \,
\phi(x_1) \, \phi(x_2) \, 
\frac{d^2{\hat\sigma}}{dt \, du}
\nonumber \\ &&
\eeqa
where  $\phi$ denotes the pdf;  
$Y$ is the top-quark rapidity,
$Y^{\pm}=\pm (1/2) \ln[(1+\beta_T)/(1-\beta_T)]$, with
$\beta_T=[1-4(m_3^2+p_T^2)S/(S+m_3^2-m_4^2)^2]^{1/2}$; 
$x_1=(s_4-m_3^2+m_4^2-x_2U_1)/(x_2 S+T_1)$,
with $T_1=T-m_3^2=-\sqrt{S} \, (m_3^2+p_T^2)^{1/2} \, e^{-Y}$ and 
$U_1=U-m_3^2=-\sqrt{S} \, (m_3^2+p_T^2)^{1/2} \, e^{Y}$; $x_2^-=(m_4^2-T)/(S+U_1)$; 
and $s_{4max}=x_2(S+U_1)+T-m_4^2$.
In particular, using Eq. (\ref{2corr}) and the properties of plus distributions, the aNNLO corrections to the $p_T$ distribution can be written as 
\beqa
\frac{d\sigma^{(2)}}{dp_T}&=& \frac{\alpha_s^2}{\pi^2} 
2\, p_T \int_{Y^-}^{Y^+} dY  
\int_{x_2^-}^1 dx_2 \, \phi(x_2) 
\nonumber \\ && \hspace{-20mm}
\times \left\{\int_0^{s_{4max}} ds_4 
\sum_{k=0}^3\frac{1}{s_4} \ln^k\left(\frac{s_4}{m_t^2}\right) 
\left[F_{\rm LO} \, C_k^{(2)} \frac{x_1 x_2 S}{x_2 S+T_1} \phi(x_1)
-F_{\rm LO}^{\rm el} \, C_k^{(2) \rm el}\frac{x_1^{\rm el} x_2 S}{x_2 S+T_1} 
\phi\left(x_1^{\rm el}\right)\right] \right.
\nonumber \\ && \hspace{-17mm} \left. 
{}+\sum_{k=0}^3 \frac{1}{k+1} \ln^{k+1}\left(\frac{s_{4max}}{m_t^2}\right) 
F_{\rm LO}^{\rm el} \, C_k^{(2) \rm el}  
\frac{x_1^{\rm el} x_2 S}{x_2 S+T_1}
\phi\left(x_1^{\rm el}\right)  \right\} \, .
\eeqa
Here the elastic versions of $x_1$, $F_{\rm LO}$, and $C_k^{(2)}$, denoted by the 
superscript ``el'', refer to these variables calculated with the constraint $s_4=0$.
We note that the total cross section can be obtained by integrating 
the $p_T$ distribution from 0 to  $p_{T max}=[(S-m_3^2-m_4^2)^2-4m_3^2m_4^2]^{1/2}/(2\sqrt{S})$,
and we have checked for consistency that we find the total cross section 
results of \cite{NK}, which are also in excellent agreement with LHC and 
Tevatron data in all three channels (see Ref. \cite{NKproc} for comparisons 
with recent data).  

\mysection{$t$-channel $p_T$ distributions}

We begin with $t$-channel single-top production. The total cross section at 13 TeV energy at the LHC for a top-quark mass $m_t=173.3$ GeV is $136^{+3}_{-1} \pm 3$ pb for single-top production and $82^{+2}_{-1} \pm 2$ pb for single-antitop production in the $t$-channel using MSTW 2008 NNLO pdf \cite{MSTW}. The central results are with $\mu_F=\mu_R=m_t$. The theoretical uncertainty of the cross section consists of two parts: the first one is from scale variation by a factor of two (i.e. from $m_t/2$ to $2 m_t$); and the second one is from the MSTW pdf \cite{MSTW} uncertainties at 90\% C.L. It is seen that the pdf uncertainties are somewhat larger than the scale uncertainties.

We note that the difference is very small if instead we use MMHT 2014 NNLO pdf \cite{MMHT}; in that case we find  $138^{+3}_{-1} \pm 2$ pb for single-top production and $83^{+2}_{-1} \pm 1$ pb for single-antitop production, where the  first uncertainty is from scale variation and the second uncertainty is from the MMHT pdf at 68\% C.L. We observe that although the central values and scale uncertainties are very close whether one uses MSTW or MMHT pdf, the pdf uncertainties in the two cases are quite different because they are 90\% C.L. (i.e. more conservative) for the former and 68\% C.L. for the latter. In the figures below we use MSTW 2008 pdf unless otherwise indicated.

\begin{figure}
\begin{center}
\includegraphics[width=8cm]{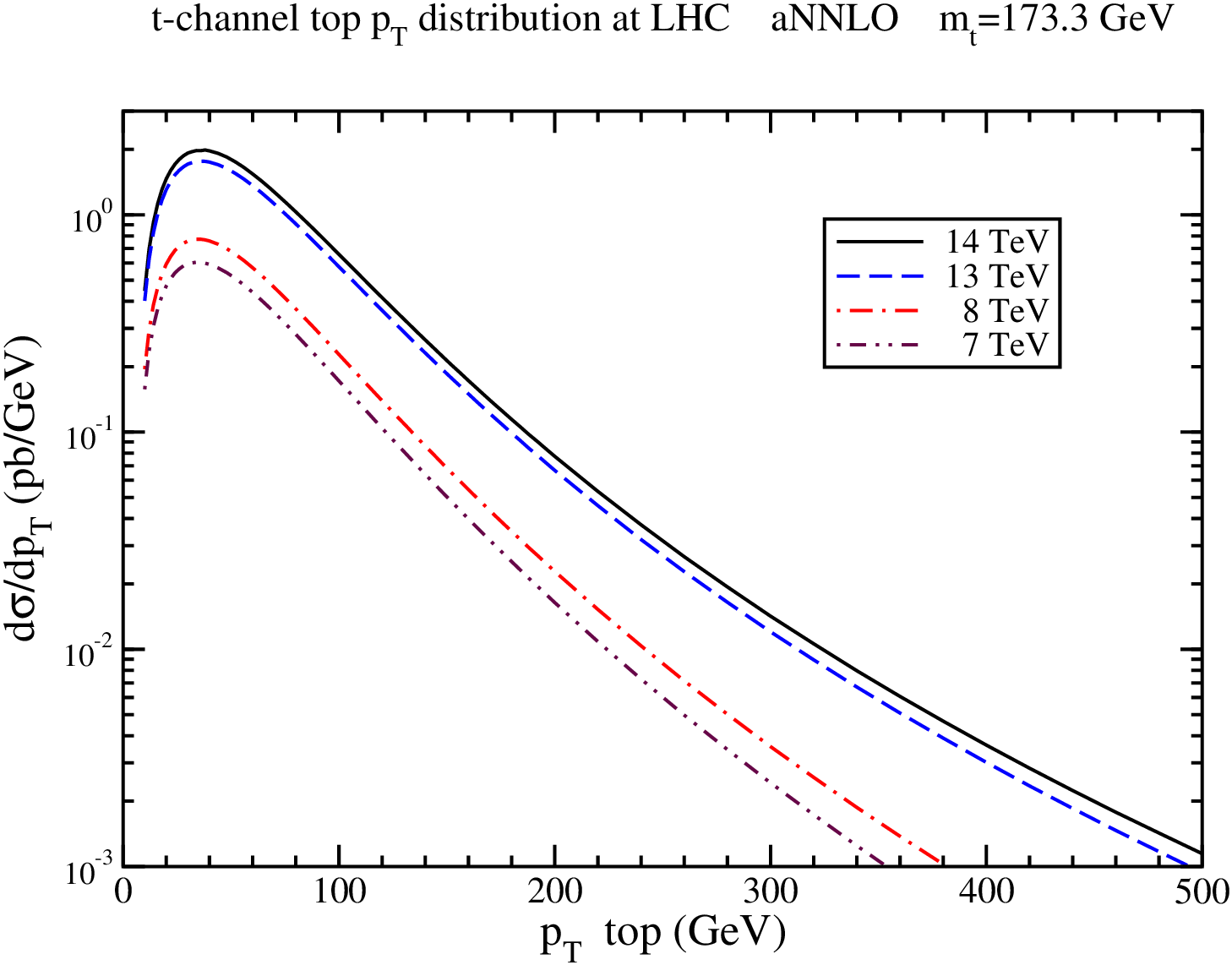}
\hspace{3mm}
\includegraphics[width=8cm]{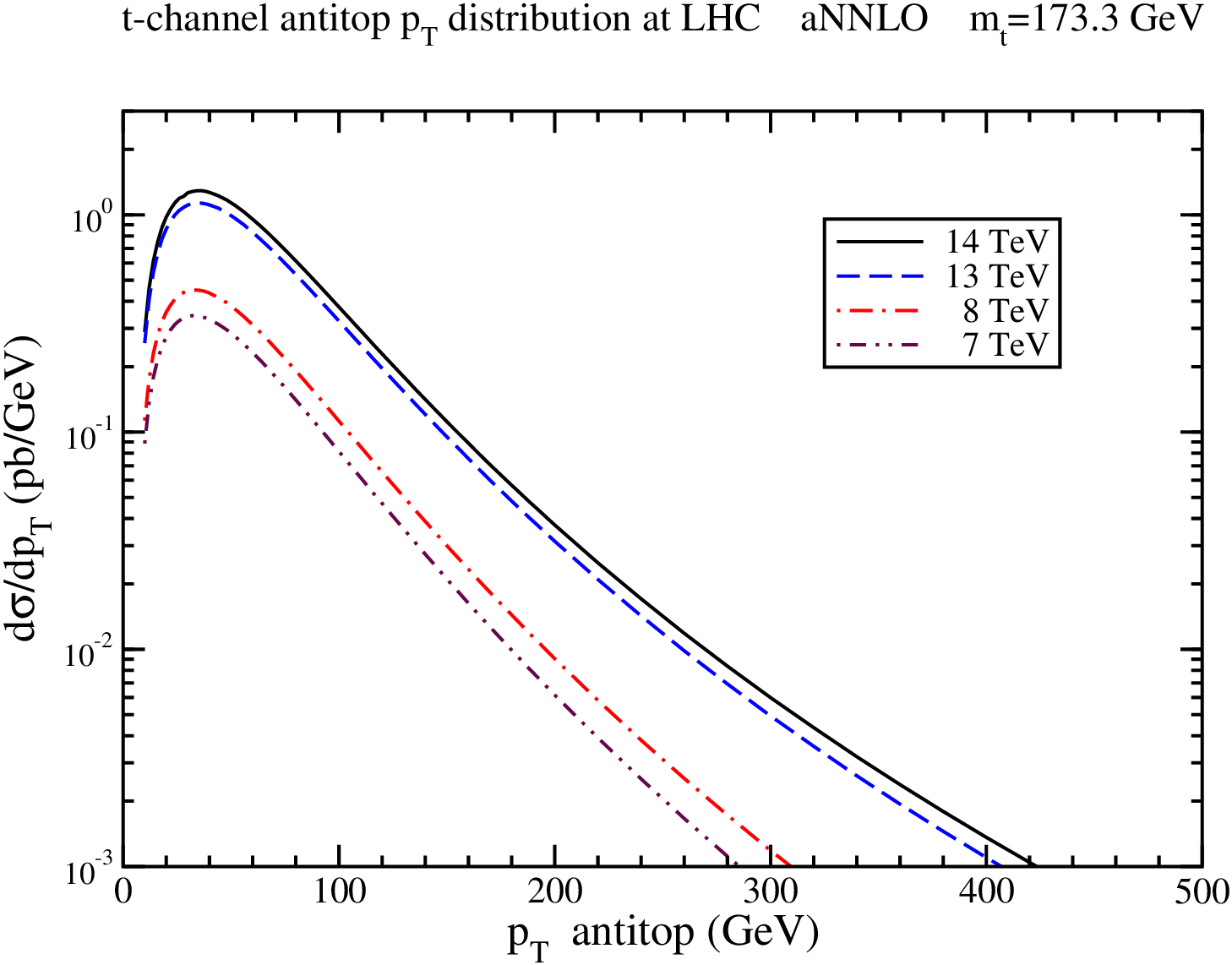}
\caption{Approximate NNLO top (left) and antitop (right) $t$-channel  
$p_T$ distributions at 7, 8, 13, and 14 TeV LHC energy.}
\label{pttchlhcplot}
\end{center}
\end{figure}

In Fig. \ref{pttchlhcplot} we present at 7, 8, 13, and 14 TeV LHC energy the central aNNLO results in the $t$-channel for the top-quark $p_T$ distribution in the left plot as well as for the antitop $p_T$ distribution in the right plot. 
The $p_T$ range displayed is up to 500 GeV and the vertical logarithmic 
scales in the two plots are chosen the same for 
ease of comparison of the relative magnitude of the distributions. 

\begin{figure}
\begin{center}
\includegraphics[width=8cm]{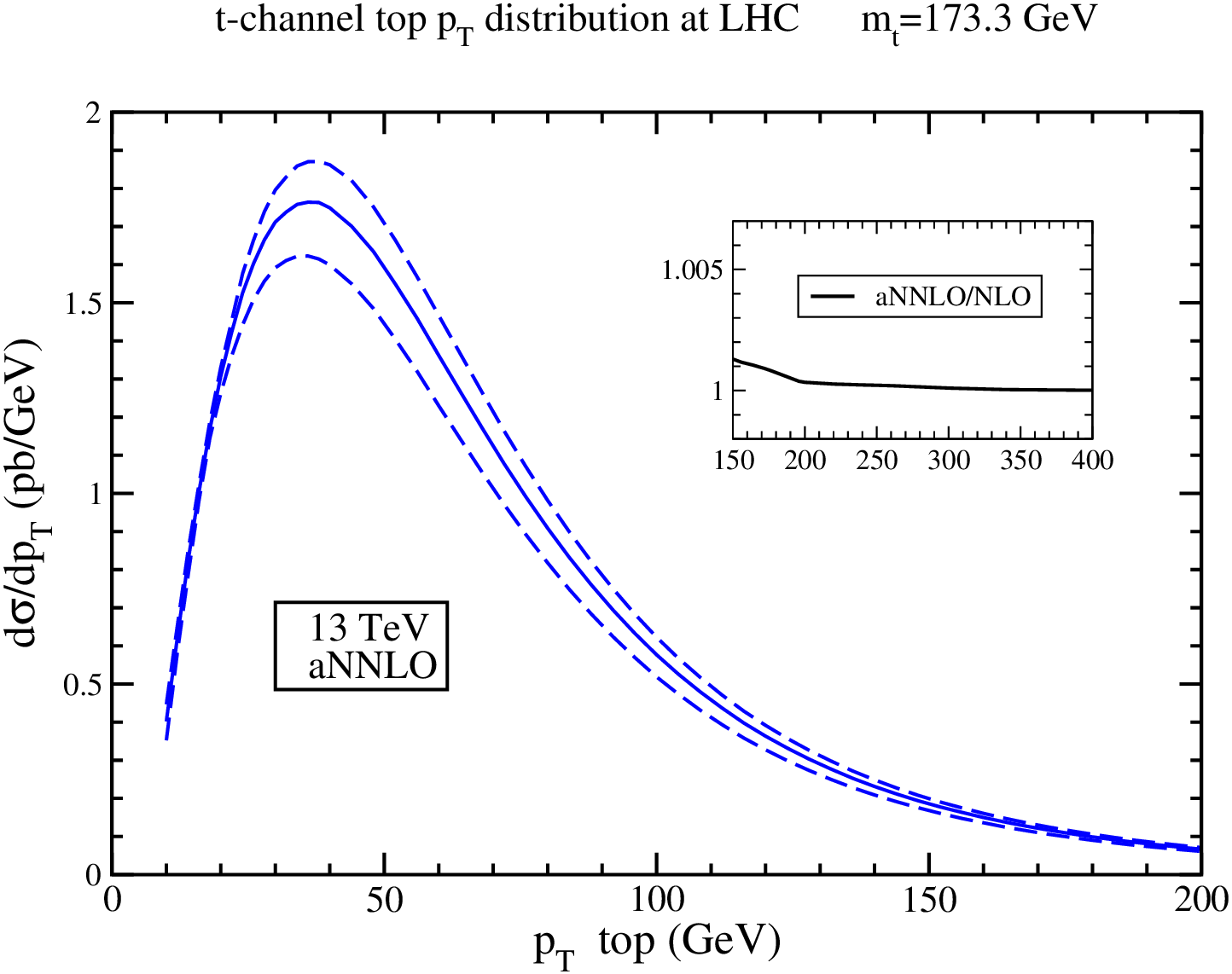}
\hspace{3mm}
\includegraphics[width=8cm]{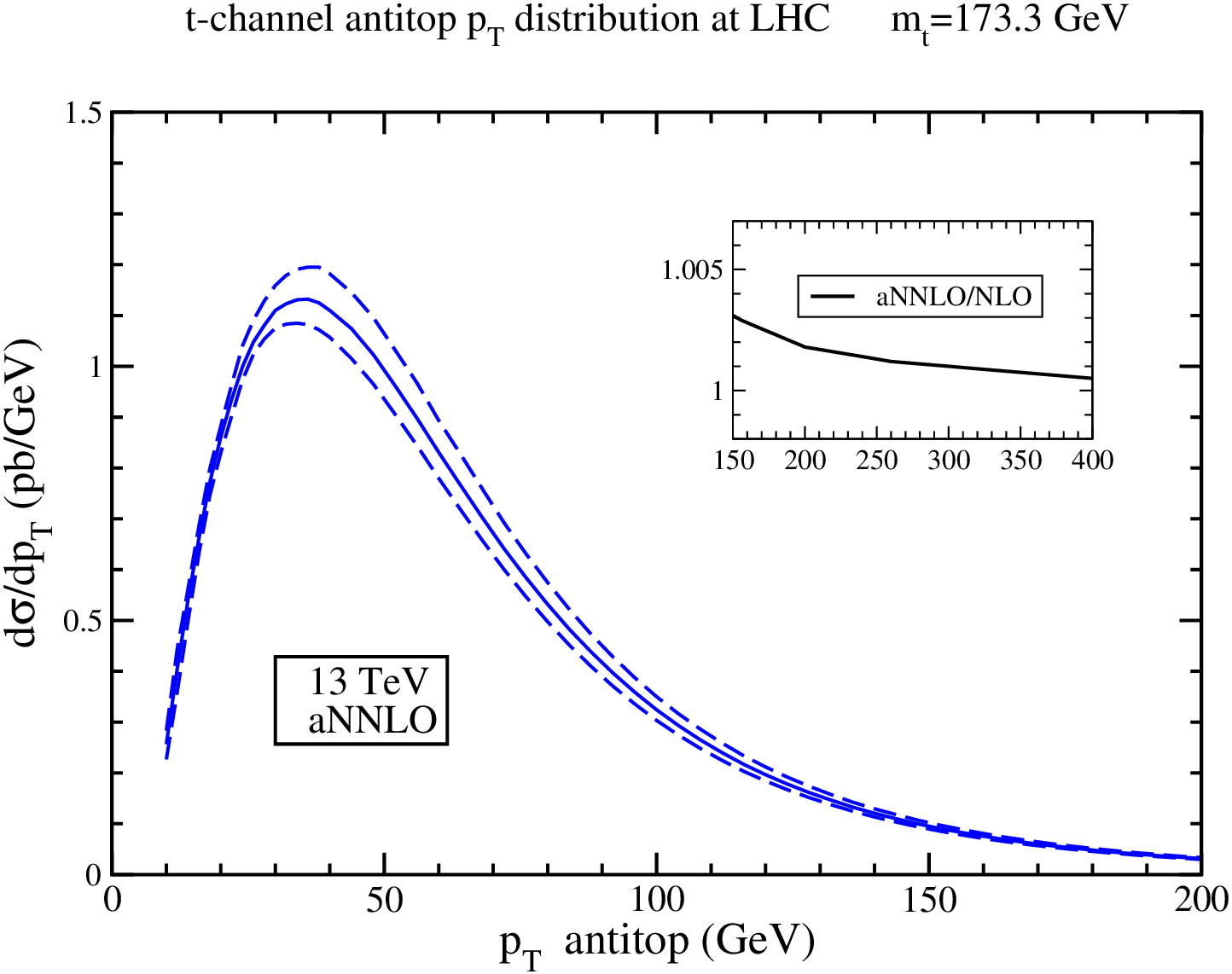}
\caption{Approximate NNLO top (left) and antitop (right) $t$-channel 
$p_T$ distributions at 13 TeV LHC energy with theoretical
uncertainty displayed by the dashed lines.}
\label{pttch13lhcplot}
\end{center}
\end{figure}

In Fig. \ref{pttch13lhcplot} we present linear plots for the aNNLO $p_T$ distribution for the top (left) and the antitop (right) in $t$-channel production at 13 TeV LHC energy. We also show the theoretical uncertainty by providing upper and lower values (dashed lines). As we noted for the total cross section, the majority of the uncertainty is due to the pdf. The top $p_T$ distributions peak at a $p_T$ of around 36 GeV, and the aNNLO corrections provide a small enhancement of 1\% over the NLO result calculated with the same pdf. The inset plot shows the ratio of the aNNLO and NLO distributions at high $p_T$ values. The enhancement diminishes at large $p_T$; we note, however, that we do not perform a targeted large-$p_T$ resummation.

\begin{figure}
\begin{center}
\includegraphics[width=8cm]{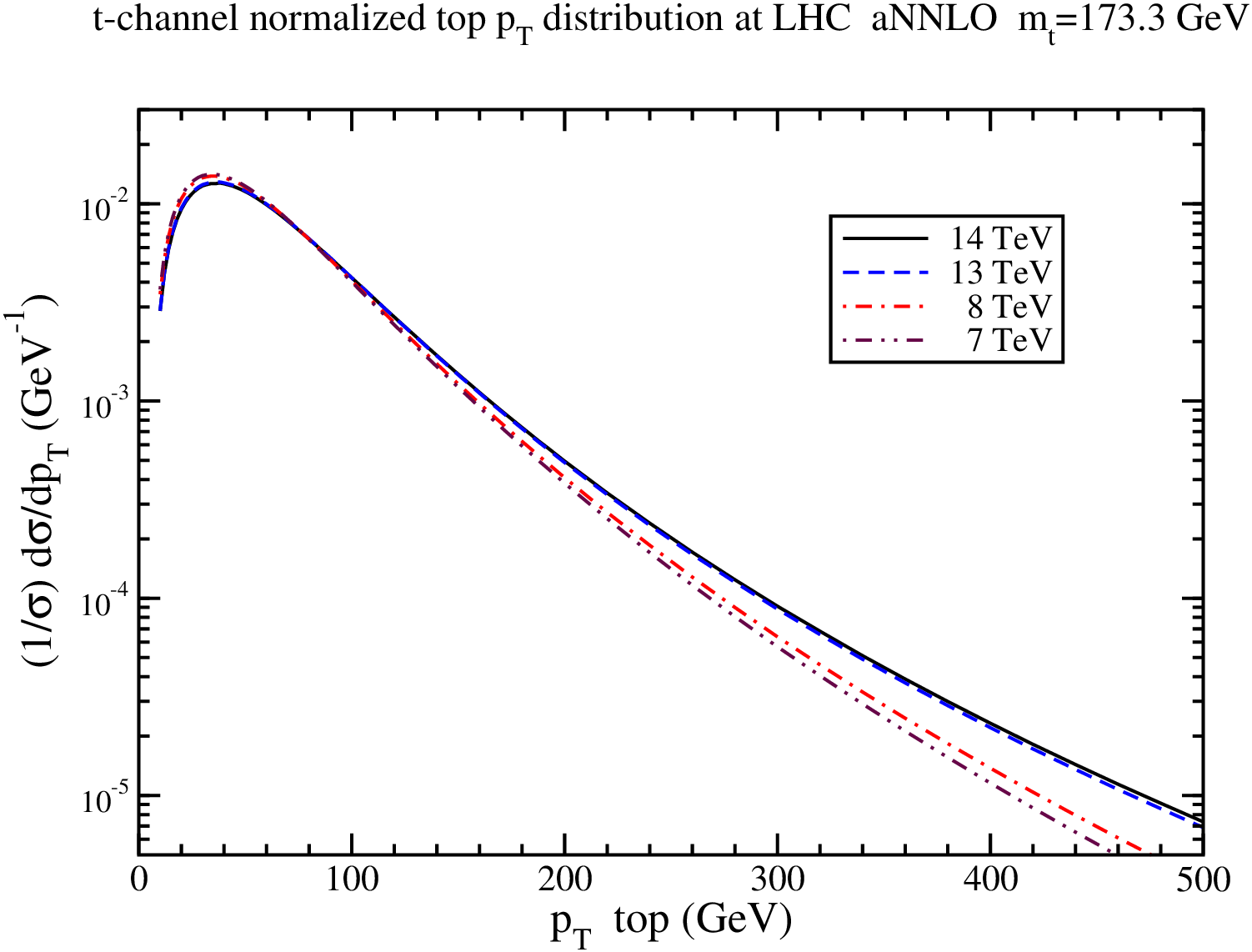}
\hspace{3mm}
\includegraphics[width=8cm]{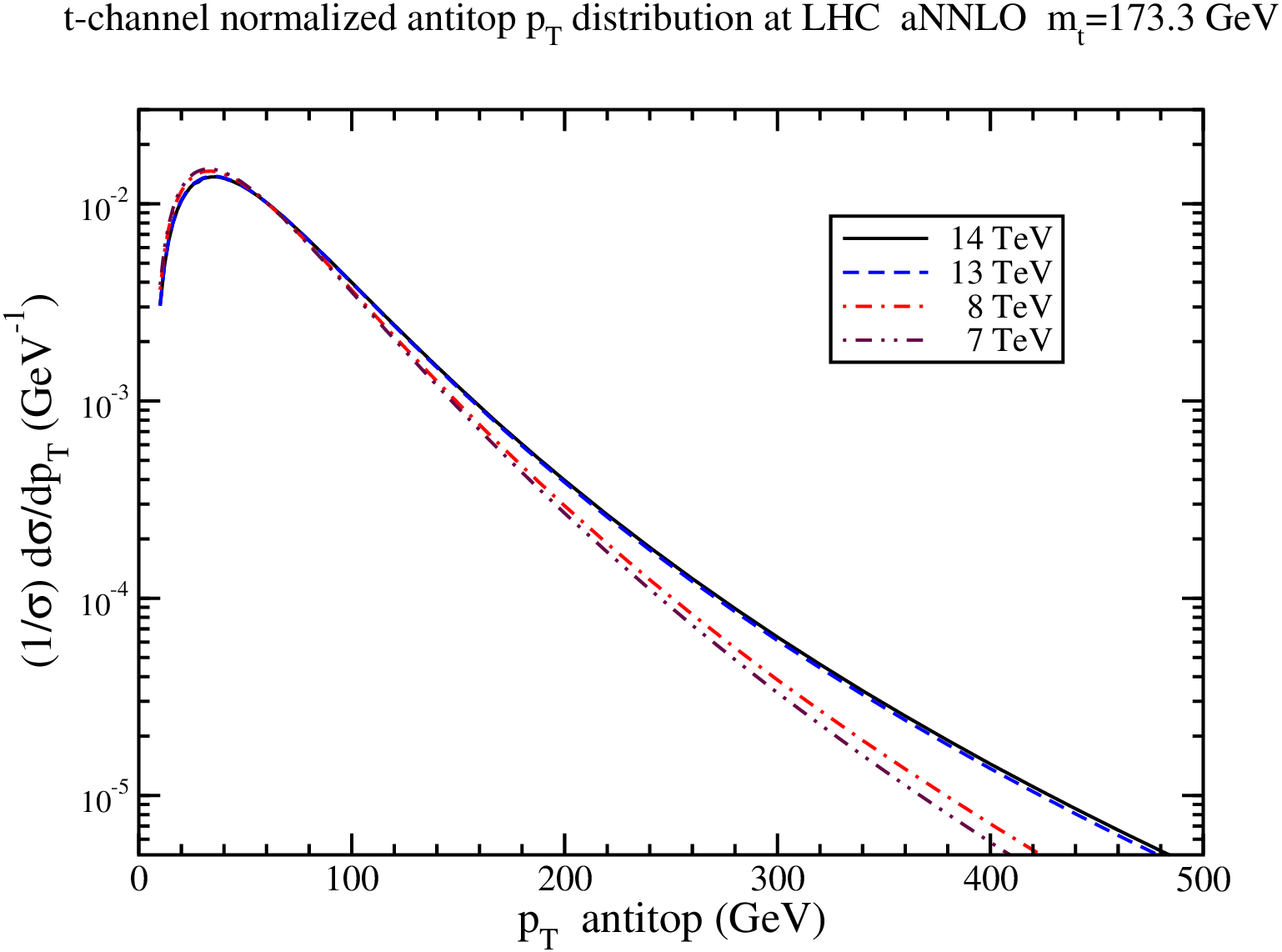}
\caption{Approximate NNLO top (left) and antitop (right) $t$-channel  
normalized $p_T$ distributions at 7, 8, 13, and 14 TeV LHC energy.}
\label{ptnormtchlhcplot}
\end{center}
\end{figure}

We note that the shape of the distributions is unaffected (to the per mille level) if MMHT 2014 pdf are instead used. There is only a very small overall normalization change as for the cross section. If one plots the normalized distribution $(1/\sigma) d\sigma/dp_T$ using the two different pdf, then the two curves are indistinguishable. In Fig. \ref{ptnormtchlhcplot} we plot the $t$-channel top (left) and antitop (right) normalized $p_T$ distributions at 7, 8, 13, and 14 TeV LHC energies.

\mysection{$s$-channel $p_T$ distributions}

We continue with $s$-channel single-top production. The total cross section at 13 TeV energy at the LHC for a top-quark mass $m_t=173.3$ GeV is $7.07 \pm 0.13 {}^{+0.24}_{-0.22}$ pb for single-top production and $4.10 \pm 0.05 {}^{+0.14}_{-0.16}$ pb for single-antitop production in the $s$-channel. As before, the theoretical uncertainty consists of two parts: the first one is from scale variation by a factor of two, and the second and larger one is from the MSTW pdf \cite{MSTW} 90\% C.L. uncertainties.  Again, we note that the difference is very small if instead we use MMHT 2014 NNLO pdf \cite{MMHT}; in that case we find $7.15 \pm 0.13 {}^{+0.15}_{-0.17}$ pb for single-top production and $4.14 \pm 0.05 \pm 0.10$ pb for single-antitop production, where the pdf uncertainty is at 68\% C.L.

In the $s$-channel, the enhancement from the NNLO soft-gluon corrections is significant, in contrast to the $t$-channel. We find an enhancement of over 8\% for the total aNNLO $s$-channel cross section relative to NLO. 

\begin{figure}
\begin{center}
\includegraphics[width=8cm]{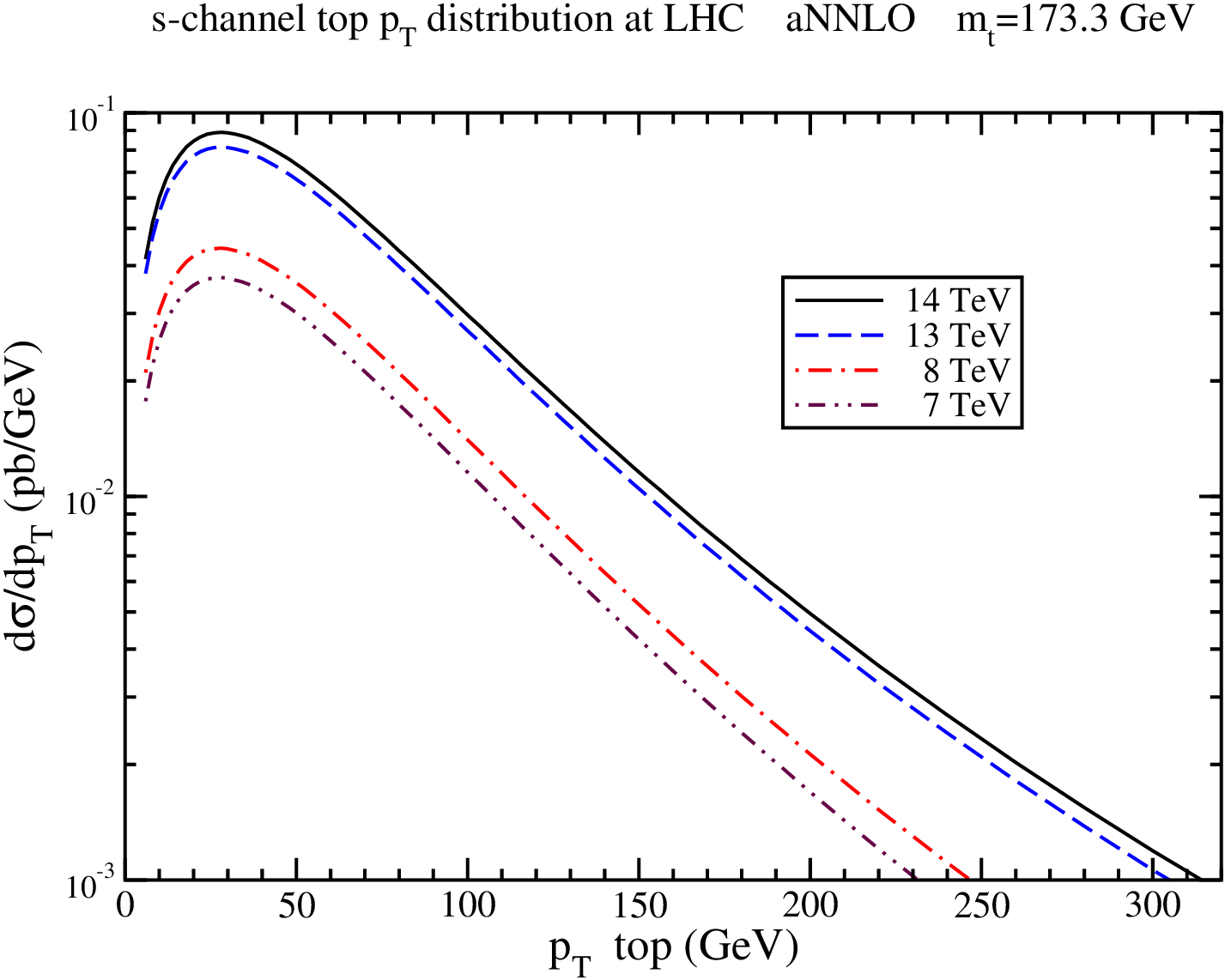}
\hspace{3mm}
\includegraphics[width=8cm]{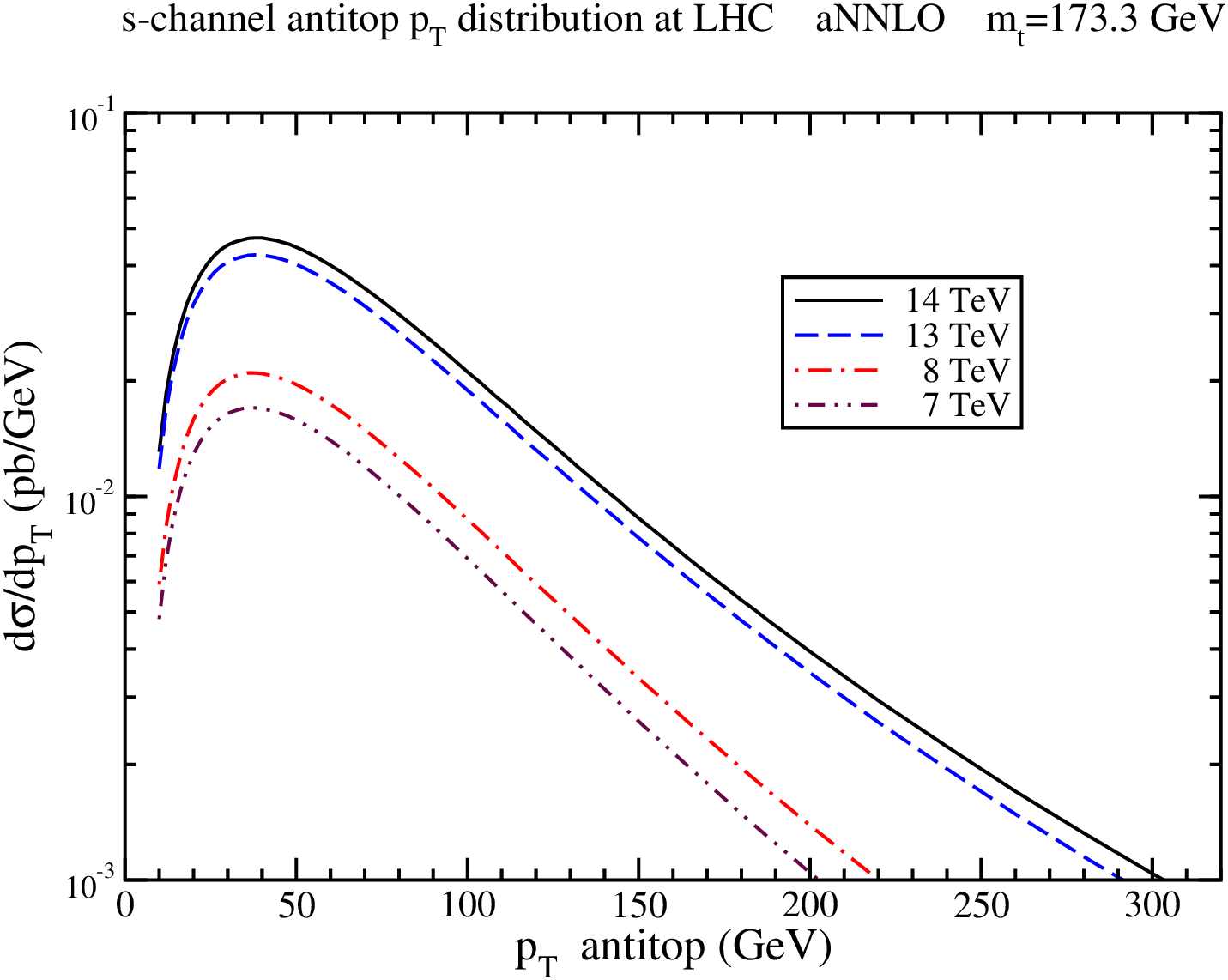}
\caption{Approximate NNLO top (left) and antitop (right)  
$s$-channel $p_T$ distributions at 7, 8, 13, and 14 TeV LHC energy.}
\label{ptschlhcplot}
\end{center}
\end{figure}

In Fig. \ref{ptschlhcplot} we present the $s$-channel central aNNLO results 
for the top-quark $p_T$ distribution in the left plot as well as for the 
antitop $p_T$ distribution in the right plot at 7, 8, 13, and 14 TeV LHC 
energy. The $p_T$ range displayed is up to 320 GeV and the vertical logarithmic 
scales in the two plots are again chosen to be identical. 

\begin{figure}
\begin{center}
\includegraphics[width=8cm]{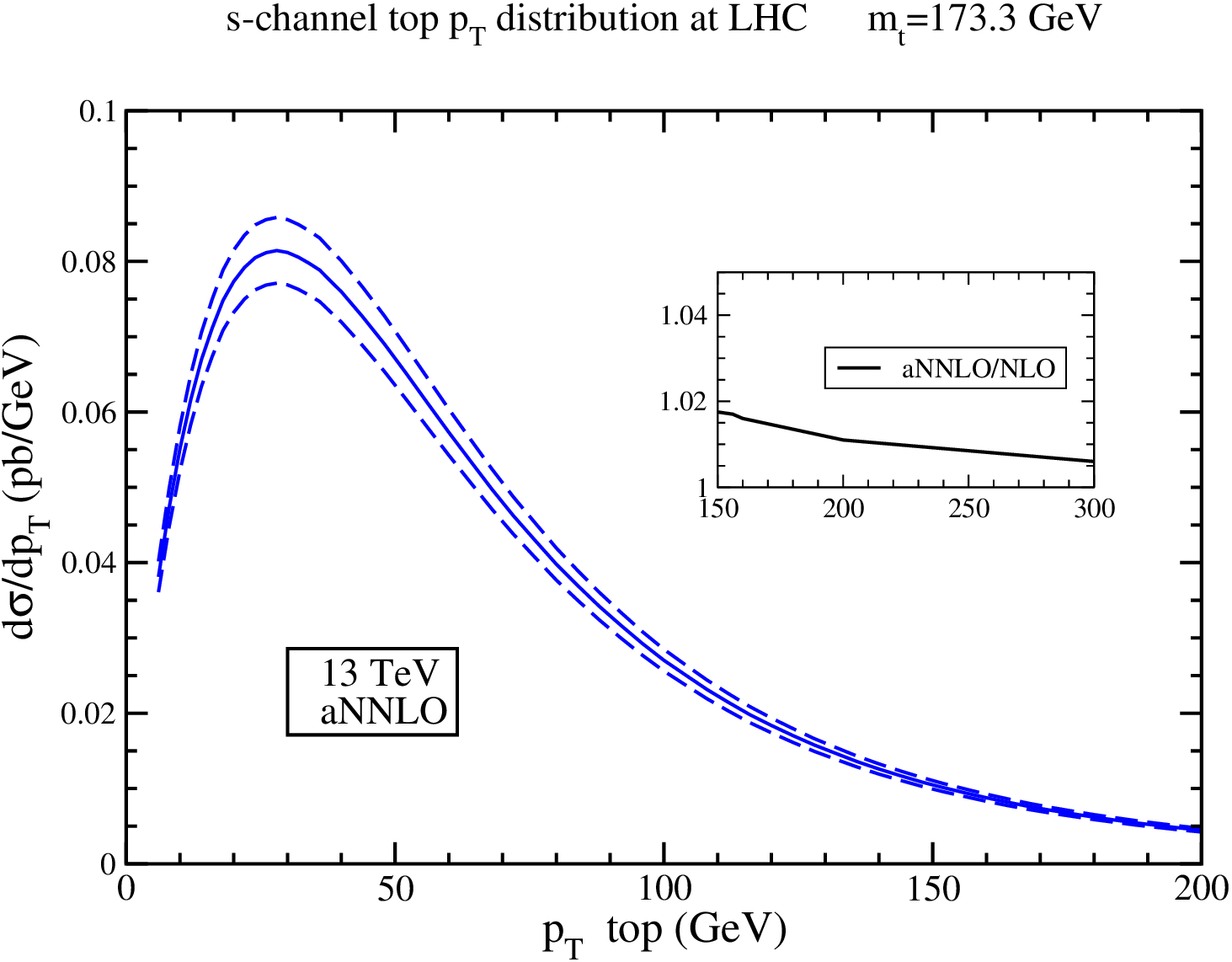}
\hspace{3mm}
\includegraphics[width=8cm]{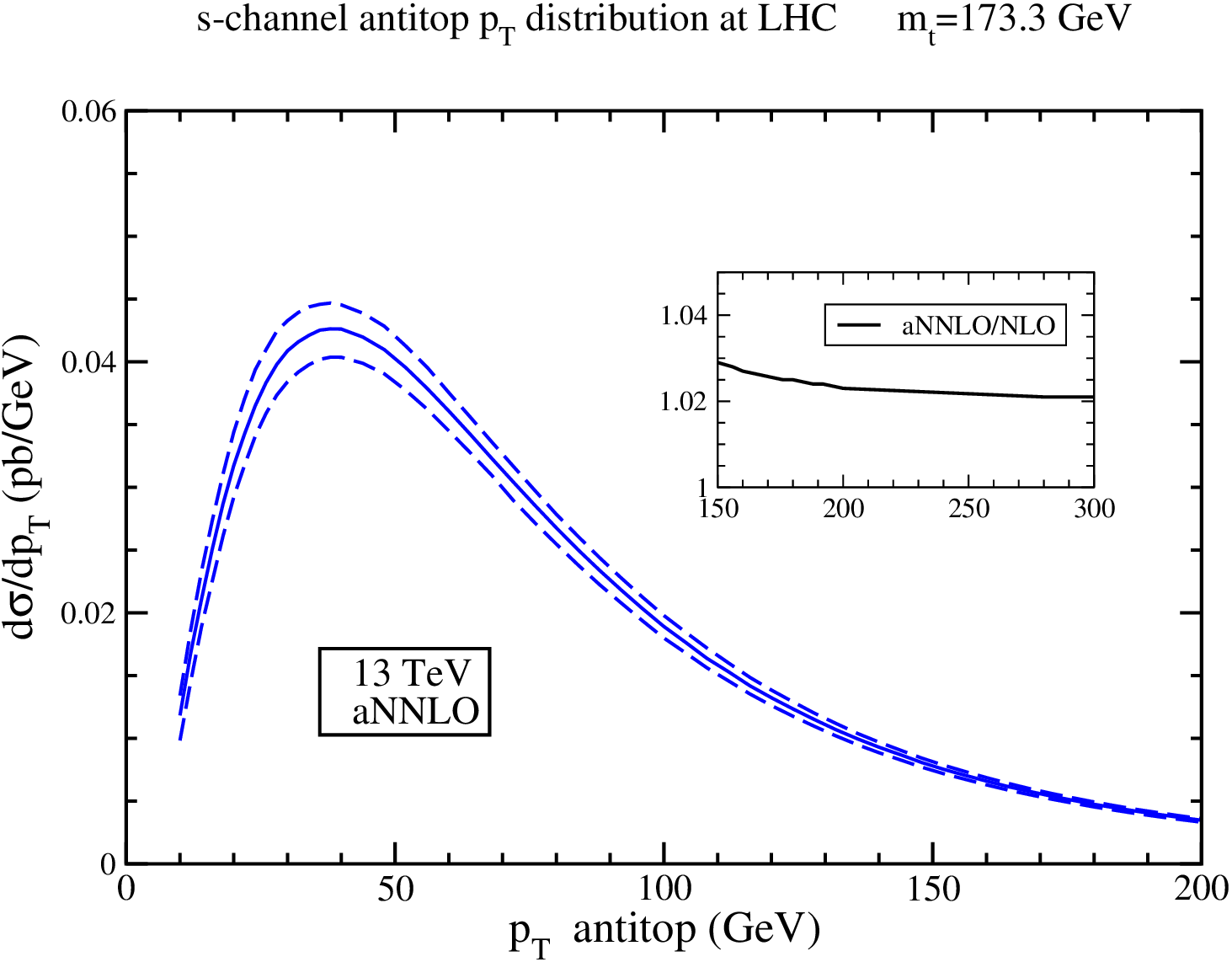}
\caption{Approximate NNLO top (left) and antitop (right) $s$-channel 
$p_T$ distributions at 13 TeV LHC energy with theoretical uncertainty 
displayed by the dashed lines.}
\label{ptsch13lhcplot}
\end{center}
\end{figure}

In Fig. \ref{ptsch13lhcplot} we present linear plots for the aNNLO $p_T$ distribution for the top (left) and the antitop (right) in $s$-channel production at 13 TeV LHC energy. As before, we show the theoretical uncertainty by providing upper and lower values. The top $p_T$ distributions peak at a $p_T$ of around 28 GeV, and the aNNLO corrections provide a large enhancement over the NLO result. The inset plot shows the ratio aNNLO/NLO at high $p_T$ where the enhancement is smaller.

\begin{figure}
\begin{center}
\includegraphics[width=8cm]{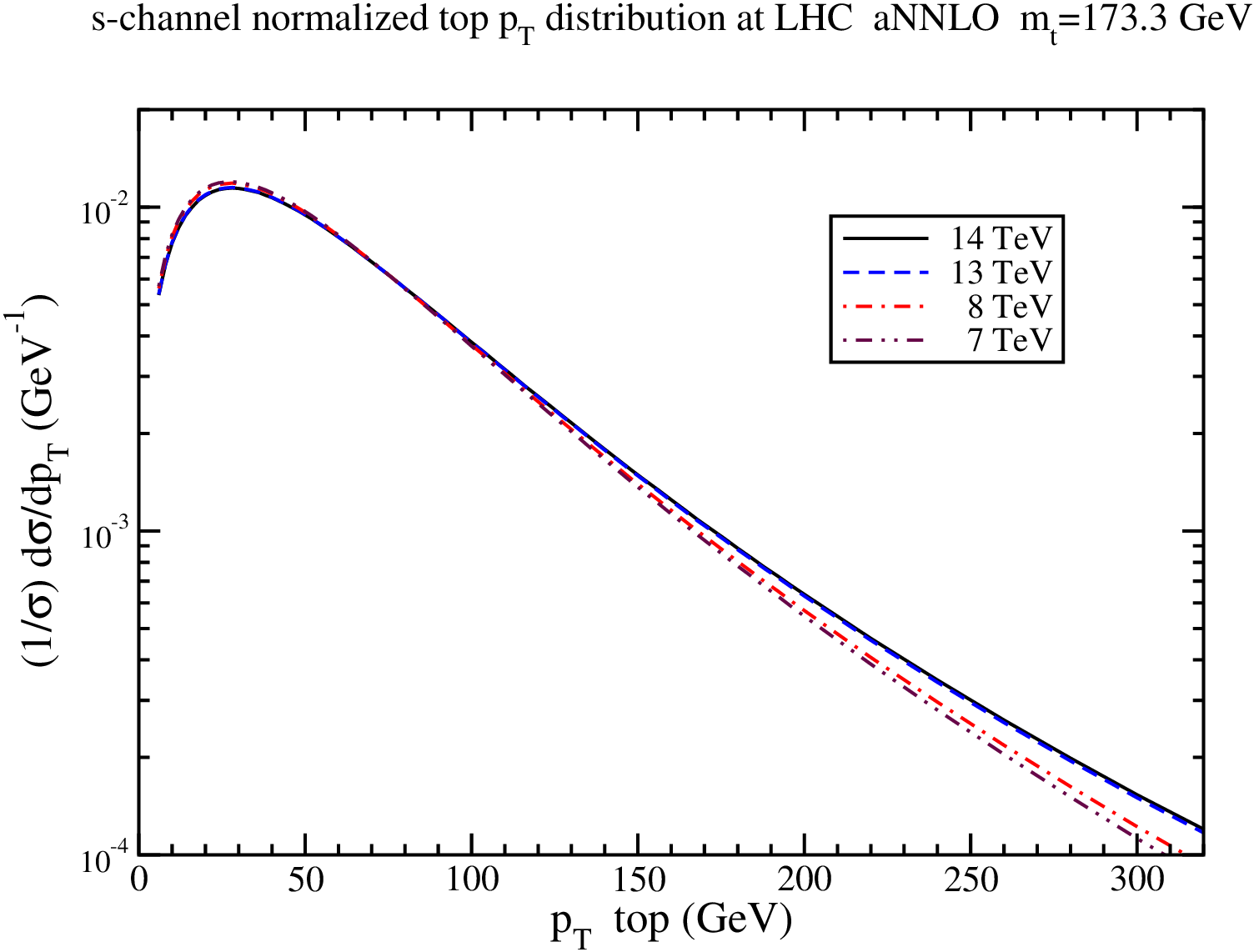}
\hspace{3mm}
\includegraphics[width=8cm]{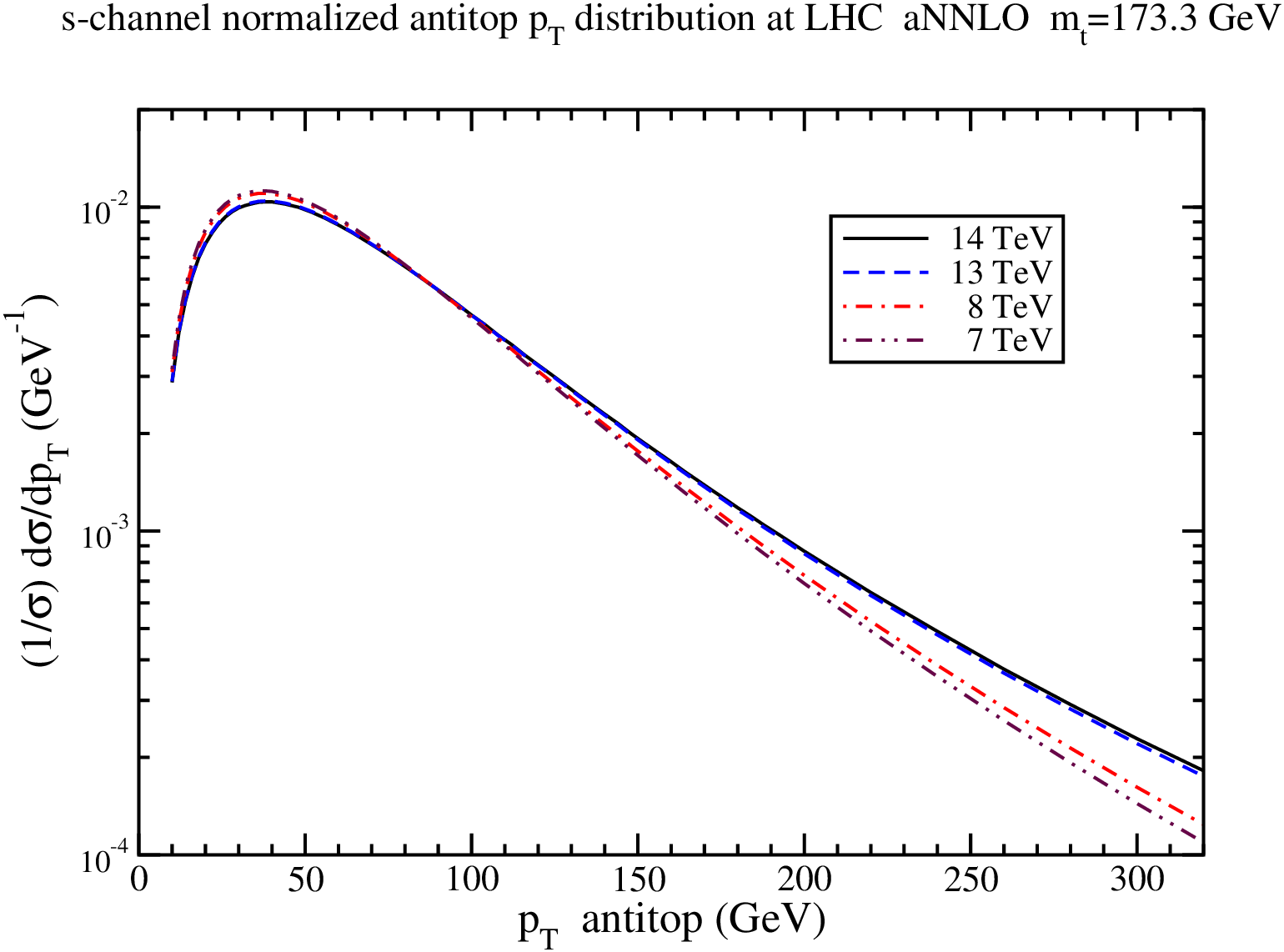}
\caption{Approximate NNLO top (left) and antitop (right)  
normalized $s$-channel $p_T$ distributions at 7, 8, 13, and 14 TeV LHC energy.}
\label{ptnormschlhcplot}
\end{center}
\end{figure}

Again, we note that the shape of the distributions is unaffected if MMHT 2014 pdf are instead used. In Fig. \ref{ptnormschlhcplot} we plot the $s$-channel top (left) and antitop (right) normalized $p_T$ distributions,  $(1/\sigma) d\sigma/dp_T$, at 7, 8, 13, and 14 TeV LHC energies.

\mysection{$tW$-channel $p_T$ distributions}

Finally, we discuss $tW$ production. The total cross section at 13 TeV energy at the LHC for a top quark mass $m_t=173.3$ GeV using MSTW 2008 pdf \cite{MSTW} is $35.2 \pm 0.9 {}^{+1.6}_{-1.7}$ pb for $tW^-$ production, and it is the same for ${\bar t} W^+$ production. Again, the theoretical uncertainty comes from scale variation by a factor of two, and from the 90\% C.L. pdf uncertainty; the latter is almost a factor of two larger. Again, we note that the difference is very small if instead we use MMHT 2014 NNLO pdf \cite{MMHT}; in that case we find  $36.3 \pm 0.9 \pm 0.9$ pb, where again the pdf uncertainty is at 68\% C.L.

The enhancement from the NNLO soft-gluon corrections is also large in the $tW$ channel. We find an 8\% increase for the total aNNLO $tW$ cross section relative to NLO. 

\begin{figure}
\begin{center}
\includegraphics[width=8cm]{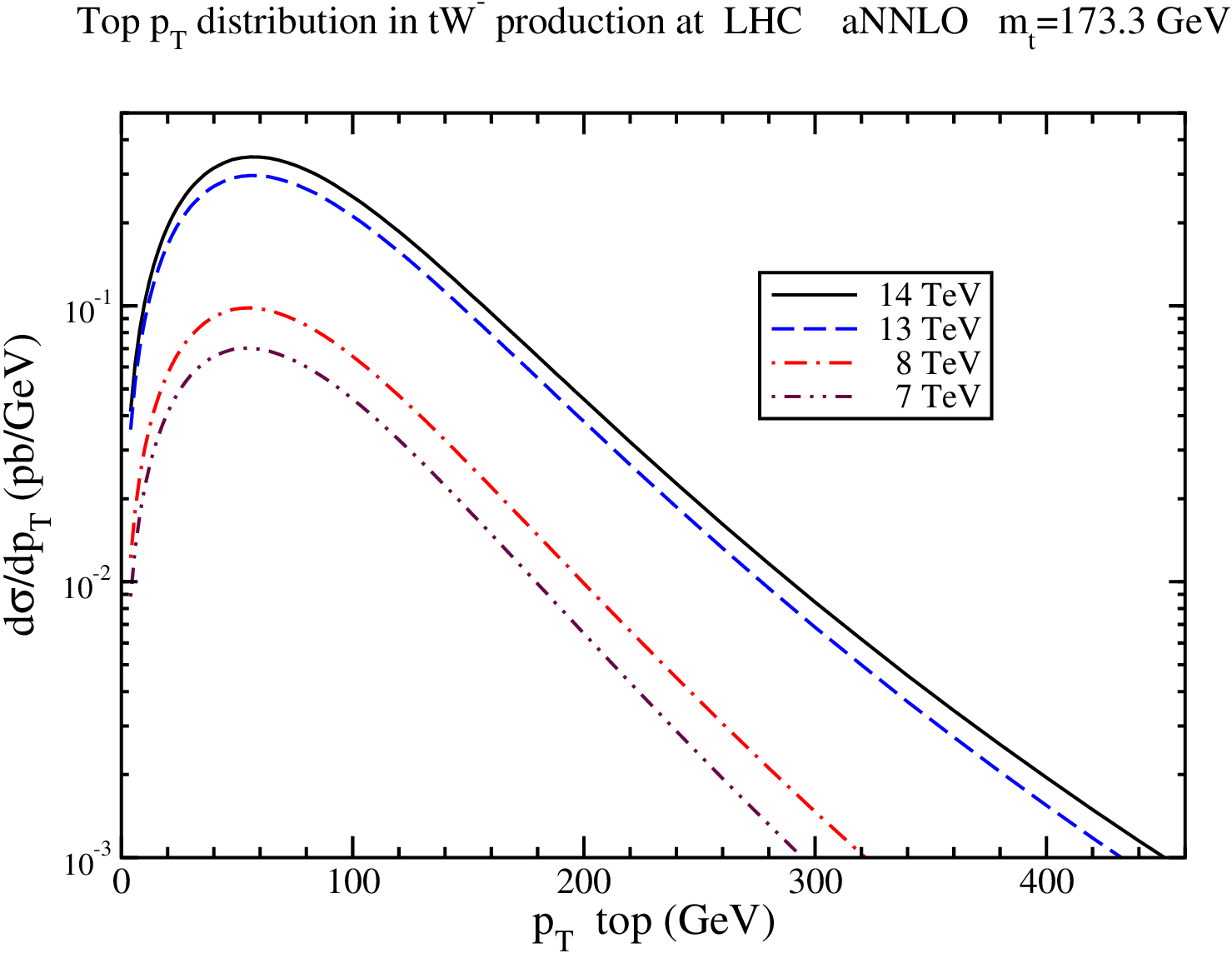}
\hspace{3mm}
\includegraphics[width=8cm]{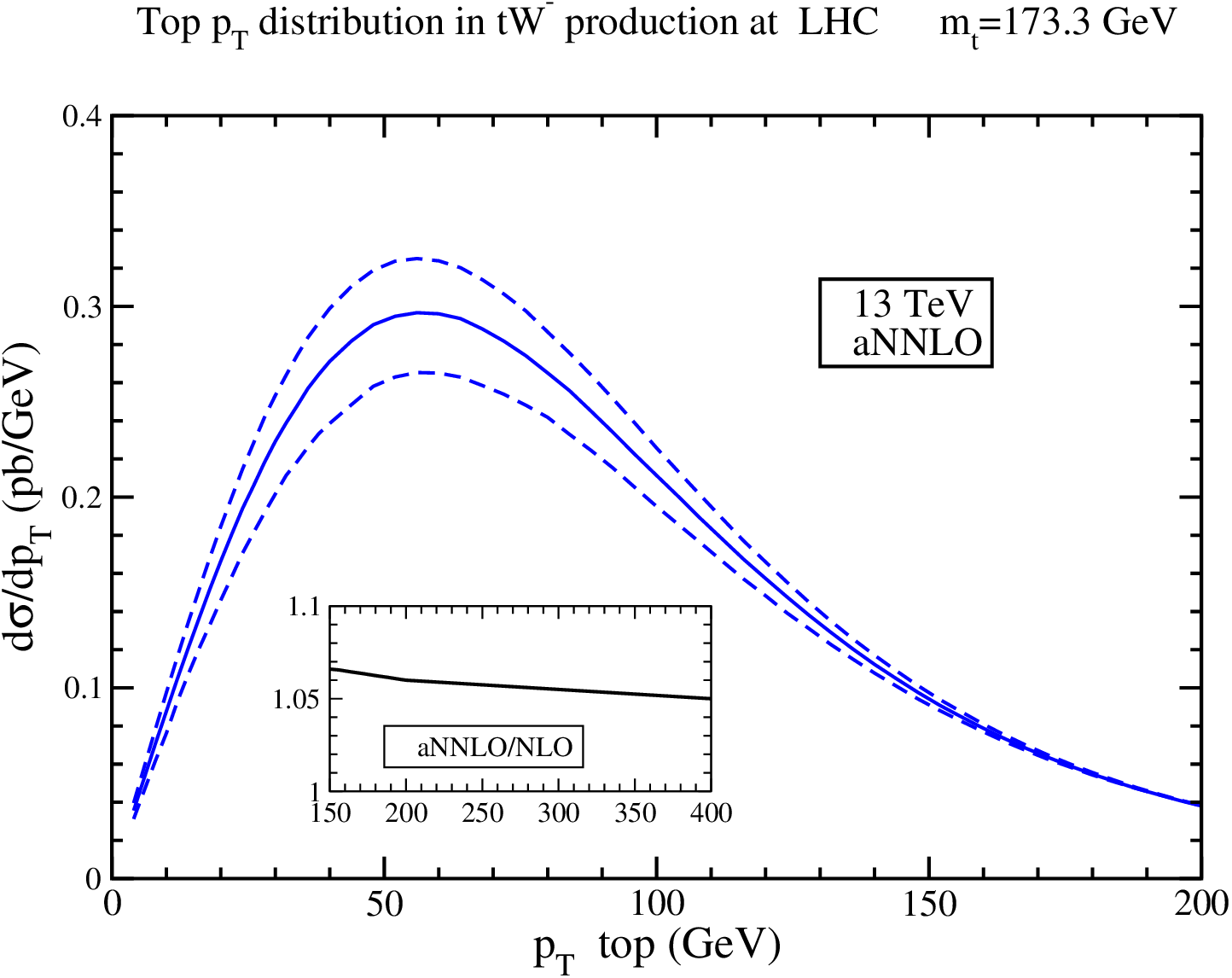}
\caption{Approximate NNLO top-quark $p_T$ distributions in the $tW$ channel 
at (left) 7, 8, 13, and 14 TeV LHC energy, and (right) at 13 TeV with theoretical uncertainty displayed.}
\label{pttWlhcplot}
\end{center}
\end{figure}

In the left plot of Fig. \ref{pttWlhcplot} we present the central aNNLO results for the top-quark $p_T$ distribution in $tW^-$ production at 7, 8, 13, and 14 TeV LHC energy. In the right plot of Fig. \ref{pttWlhcplot} we present a linear plot for the aNNLO top $p_T$ distribution in $tW^-$ production at 13 TeV LHC energy. We also show the theoretical uncertainty by providing upper and lower values. The top $p_T$ distributions peak at a $p_T$ of around 56 GeV, and the aNNLO corrections provide a substantial enhancement of 8.5\% over the NLO result. The inset plot shows the ratio aNNLO/NLO at high $p_T$. The $p_T$ distributions for the antitop in this channel are the same as for the top.

\begin{figure}
\begin{center}
\includegraphics[width=8cm]{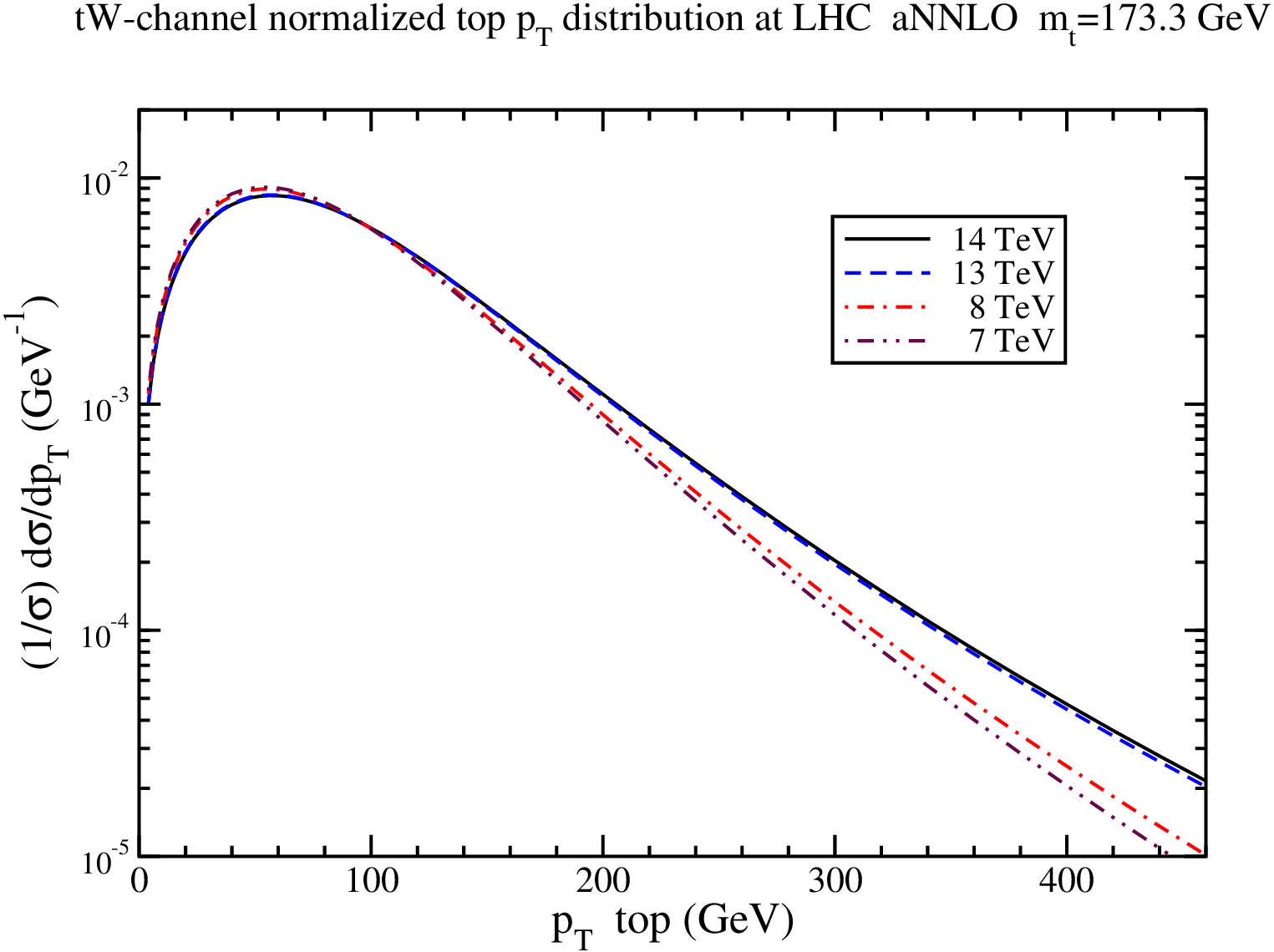}
\caption{Approximate NNLO $tW$-channel normalized top $p_T$ distributions at 7, 8, 13, and 14 TeV LHC energy.}
\label{ptnormtWlhcplot}
\end{center}
\end{figure}

Once again, we note that the shape of the distributions is unaffected if MMHT 2014 pdf are instead used. In Fig. \ref{ptnormtWlhcplot} we plot the $tW$-channel normalized top $p_T$ distributions,  $(1/\sigma) d\sigma/dp_T$, at 7, 8, 13, and 14 TeV LHC energies.

\mysection{Conclusions}

I have presented the single-top and single-antitop transverse-momentum distributions at approximate NNLO by including soft-gluon corrections derived from NNLL resummation. Results were presented at 7, 8, 13, and 14 LHC energies for $t$-channel, $s$-channel and $tW$ production. We have paid particular attention to the current 13 TeV LHC energy and also provided theoretical uncertainties. The corrections are large and very significant in $s$-channel and $tW$ production but they are rather small in $t$-channel production.

\mysection*{Acknowledgements}
This material is based upon work supported by the National Science Foundation 
under Grant No. PHY 1519606.

\end{document}